%% file: ising_cortex.tex
\documentclass{llncs}
\usepackage{graphicx}      
\usepackage{amsmath} 
\usepackage{amssymb} 
\usepackage{bm}
\usepackage{epstopdf}
\usepackage{amsmath,amsfonts,paralist}
\usepackage{amssymb}
\usepackage{appendix}
\usepackage{mathbbol}
\usepackage{color}

\usepackage[]{algorithm2e}

\DeclareMathOperator*{\argmax}{\arg\!\max}

\DeclareMathOperator{\bme}{ \bm{\eta} }
\DeclareMathOperator{\bmn}{ \bm{\nabla} }
\DeclareMathOperator{\bmTO}{ \bf{T}_\Omega }
\DeclareMathOperator{\bmTE}{ \bf{T}_\eta }
\DeclareMathOperator{\bmTEMC}{ \bf{T}_\eta^\text{MC} }
\DeclareMathOperator{\bmnlMC}{ {\bf \nabla} \ell_{\eta}^\text{MC} }
\DeclareMathOperator{\bmnlMCt}{ {\bf \nabla} \ell_{\eta_n}^\text{MC} }
\newcommand\norm[1]{\left\lVert#1\right\rVert}

\def\to{\rightarrow}

\begin{document}

\title{Pairwise Ising model analysis of human cortical neuron recordings}

\author{Trang-Anh Nghiem\inst{1} \and Olivier Marre\inst{2} Alain Destexhe\inst{1} \and Ulisse Ferrari\inst{2}\inst{3} } 
\tocauthor{Trang-Anh Nghiem, Olivier Marre, Alain Destexhe, Ulisse Ferrari} 
 \institute{Laboratory of Computational Neuroscience, Unit\'{e} de Neurosciences, Information et 
Complexit\'{e}, CNRS, Gif-Sur-Yvette, France. \and Institut de la Vision, INSERM and UMPC, 17 rue Moreau, 75012 Paris, France. \and ulisse.ferrari@gmail.com }

\maketitle              

\begin{abstract}
 \input{abstract}

\end{abstract}  

\input{introduction}

\section{The model and the geometry of the parameter space}
\input{model}

\section{Inference algorithm}
\input{inference}

\section{Analysis of cortical recording}
\input{analysis}

\vspace{0.5cm}
\textbf{Acknowledgments $\quad$}
We thank B. Telenczuk, G. Tkacik and M. Di Volo  for useful discussion.
Research funded by European Community (Human Brain Project, H2020-720270),  ANR TRAJECTORY, ANR OPTIMA, French State program
Investissements d’Avenir managed by the Agence Nationale de la Recherche  [LIFESENSES: ANR-10-LABX-65] and NIH grant U01NS09050.




%
%
%


\end{document}

%% file: abstract.tex
During wakefulness and deep sleep brain states, cortical neural networks show a different behavior, with the second characterized by transients of high network activity.
To investigate their impact on neuronal behavior, we apply a pairwise Ising model analysis by inferring the maximum entropy model that reproduces single and pairwise moments of the neuron's spiking activity.
In this work we first review the inference algorithm introduced in Ferrari, \textit{Phys. Rev. E} (2016) \cite{Ferrari16a}.
We then succeed in applying the algorithm to infer the model from a large ensemble of neurons recorded by multi-electrode array in human temporal cortex.
We  compare the Ising model performance in capturing the statistical properties of the network activity during wakefulness and deep sleep.
For the latter, the pairwise model misses relevant transients of high network activity, suggesting that additional constraints are necessary to accurately model the data.

%% file: introduction.tex
Advances in experimental techniques have recently enabled the recording of the activity of tens to hundreds of neurons simultaneously \cite{Stevenson11} and has spurred the interest in modeling their collective behavior  \cite{Schneidman06,Peyrache12,Hamilton13,Tkacik14,Dehghani16,Gardella16,Tavoni17}.
To this purpose, the pairwise Ising model has been introduced as the maximum entropy (most generic \cite{Jaynes82}) model able to reproduce the first and second empirical moments of the recorded neurons.
Moreover it has already been applied to different brain regions in different animals \cite{Schneidman06,Hamilton13,Tkacik14,Tavoni17} and shown to work efficiently \cite{Ferrari17} .

The inference problem for a pairwise Ising model is a computationally challenging task \cite{Ackley85}, that requires devoted algorithms \cite{Broderik07,Cocco11,Sohl-Dickstein11}.
Recently, we proposed a \textit{data-driven} algorithm and applied it on rat retinal recordings \cite{Ferrari16a}.
In the present work we first review the algorithm structure and then describe our successful application to a recording in the human temporal cortex \cite{Peyrache12}.

We use the inferred Ising model to test if a model that reproduces empirical pairwise covariances without assuming any other additional information, also predicts empirical higher-order statistics.
We apply this strategy separately to brain states of wakefulness (Awake) and Slow-Wave Sleep (SWS). 
In contrast to the former, the latter is known to be characterized by transients of high activity that modulate the whole population behavior \cite{renart10}.
Consistently, we found that the Ising model does not account for such global oscillations of the network dynamics.
We do not address Rapid-Eye Movement (REM) sleep.

%% file: model.tex
The pairwise Ising model is a fully connected Boltzmann machine without hidden units.
Consequently it belongs to the exponential family and has probability distribution:
\begin{equation}
P_\eta \big(~ {\bm X}~ \big) = \exp \big(~{\bm T(X)} \cdot {\bm \eta} ~-~ \log Z[{\bm \eta}]  ~\big)~,   \label{eq:model}
\end{equation}
where  ${\bm X} \in [0,1]^N$ is the row vector of the $N$ system's free variables and $Z[{\bm \eta}]$ is the normalization.
${\bm \eta} \in \mathcal{R}^D$ is the column vector of model parameters, with $D=N(N+1)/2$ and ${\bm T(X)} \in [0,1]^D$ is the vector of model sufficient statistics.
For the fully-connected pairwise Ising model the latter is composed of the list of free variables ${\bm X}$ and their pairwise products:
\begin{equation}
  \{T_a({\bm X})\}_{a=1}^D = \{~ \{X_i\}_{i=1}^N~,   \{X_i X_j\}_{i=1,j=i+1}^N~\} \in [0,1]^D~.
\end{equation}
 A dataset $\Omega$ for the inference problem is composed by a set of $\tau_\Omega$  \textit{i.i.d.} empirical configurations ${\bm X}$:  $\Omega = \{ {\bm X}(t) \}_{t=1}^{\tau_\Omega}$. 
We cast the inference problem as a log-likelihood maximization task, which for the model (\ref{eq:model}) takes the shape:
\begin{equation}
{\bm \eta^*} \equiv \argmax_{\bm \eta} \ell [{\bm \eta}]~; \quad  \ell [{\bm \eta}] \equiv \bmTO \cdot \bme  - \log Z[{\bm \eta}] \label{eq:logL}~,
\end{equation} 
where $\bmTO \equiv \mathbf{E}\big[~ {\bm T}({\bm X})~|~ \Omega~\big]$ is the empirical mean of the sufficient statistics.
As a consequence of the exponential family properties, the log-likelihood gradient may be written as:
\begin{equation}
\bmn \ell[\bme] = \bmTO - \bmTE~,\label{eq:logLgrad}
\end{equation}
where $\bmTE = \mathbf{E}\big[~ {\bm T}({\bm X})~\big| \bme~\big]$ is the mean of ${\bm T}({\bm X})$ under the model distribution (\ref{eq:model}) with parameters $\bme$.
Maximizing the log-likelihood is then equivalent to imposing $\bmTO = \bmTE$: the inferred model then reproduces the empirical averages.

\textbf{Parameter space geometry $\quad$}
In order to characterize the geometry of the model parameter space, we define the minus log-likelihood Hessian $\mathcal{H}[\bme]$, the model Fisher matrix $J[\bme]$ and the model susceptibility matrix $\chi[\bme]$ as:
\begin{eqnarray}
\chi_{ab}[\bme] &\equiv& \mathbf{E} \big[ ~ T_a T_b ~\big| ~\bme ~\big] -  \mathbf{E} \big[ ~ T_a  ~\big| ~\bme ~\big]   \mathbf{E} \big[ ~  T_b ~\big| ~\bme ~\big] \\
J_{ab}[\bme] &\equiv&  \mathbf{E} \big[ ~   \nabla_a   \log P_\eta \big( {\bf X }  \big) ~  \nabla_b  \log P_\eta \big( {\bf X }\big) ~\big| ~\bme ~\big]  \label{fisher} ~, \\
\mathcal{H}_{ab}[\bme] &\equiv& -\nabla_a \nabla_b  l \big[\bme\big]   \label{hessian}  ~,
\end{eqnarray}
As a property inherited from the exponential family,  for the Ising model (\ref{eq:model}):
\begin{equation}
 \chi_{ab}[\bme] = J_{ab}[\bme]  = \mathcal{H}_{ab}[\bme] \label{eq:equalities}~.
\end{equation}
This last property is the keystone of the present algorithm.

Moreover, the fact that the log-likelihood Hessian can be expressed as a covariance matrix ensures its non-negativity.
Some zero Eigenvalues can be present, but they can easily be addressed by $L2$-regularization \cite{Cocco11,Ferrari16a}.
The inference problem is indeed convex and consequently the solution of (\ref{eq:logL}) exists and is unique.

%% file: inference.tex
The inference task (\ref{eq:logL}) is an hard problem because the partition function $Z[\bme]$ cannot be computed analytically.
Ref. \cite{Ferrari16a} suggests applying an approximated natural gradient method to numerically address the problem.
After an initialization of the parameters to some initial value $\bme_0$, the natural gradient \cite{Amari98,Amari98b} iteratively updates their values with:
\begin{equation}
\eta_{n+1} = \eta_n - \alpha J^{-1}[\bme_n] \cdot \bmn \ell [\eta_n] ~.
\end{equation}
For sufficiently small $\alpha$, the convexity of the problem and the positiveness of the Fisher matrix ensure the convergence of the dynamics to the solution $\bme^*$.

As computing $J[\bme_n]$ at each $n$ is computationally expensive, we use (\ref{eq:equalities}) to approximate the Fisher with an empirical estimate of the susceptibility \cite{Ferrari16a}:
\begin{equation}
J[\bme] =  \chi[\bme] \approx \chi[\bme^*] \approx \chi_\Omega \equiv \text{Cov} \big[ ~ {\bm T} ~\big| ~\Omega ~\big]~. \label{eq:DD}
\end{equation}
The first approximation becomes exact upon convergence of the dynamics, $\bme_n \to \bme^*$.
The second assumes that (i) the distribution underlying the data belongs to the family (\ref{eq:model}), and that (ii) the error in the estimate of $\chi_\Omega$, arising from the dataset's finite size, is small. 

We compute $\chi_\Omega$ of Eq. (\ref{eq:DD}) only once, and then we run the inference algorithm that performs the following approximated natural gradient:
\begin{equation}
\eta_{n+1} = \eta_n - \alpha \chi_\Omega^{-1} \cdot \bmn \ell[\eta_n] ~.\label{eq:DDupdate}
\end{equation}
\textbf{Stochastic dynamics.}\footnote{The results of this section are grounded on the repeated use of central limit theorem. See \cite{Ferrari16a} for more detail.} $\quad$
The dynamics (\ref{eq:DDupdate}) require estimating $ \bmn \ell [\eta]$ and thus of $\bmTE$ at each iteration.
This is accounted by a Metropolis Markov-Chain Monte Carlo (MC), which collects $\Gamma_\eta$, a sequence of $\tau_\Gamma$ \textit{i.i.d.} samples of the distribution (\ref{eq:model}) with parameters $\bme$ and therefore estimates:
\begin{equation}
\bmTEMC \equiv \mathbf{E}\big[~ {\bm T}({\bm X})~\big|  ~\Gamma_\eta~\big]~.
\end{equation}
This estimate itself is a random variable with mean and covariance given by:
\begin{equation}
\mathbf{E}\big[~ \bmTEMC~\big|~ \{ \Gamma_\eta \}~\big] = \bmTE~; \quad \text{Cov}\big[~ \bmTEMC~\big|~\{ \Gamma_\eta \}~\big] = \frac{J[\bme]}{\tau_\Gamma}~,
\end{equation}
where $\mathbf{E}\big[~ \cdot ~\big|~ \{ \Gamma_\eta \}~\big]$  means expectation with respect to the possible realizations $\Gamma_\eta$ of the configuration sequence. 

For $\bme$ sufficiently close to $\bme^*$, after enough iterations, this last result allows us to compute the first two moments of $\bmnlMC \equiv \bmTO - \bmTEMC$, using a second order expansion of the log-likelihood (\ref{eq:logL}):
\begin{equation}
\mathbf{E}\big[~ \bmnlMC~\big|~ \{ \Gamma_\eta \} ~\big] = \mathcal{H}[\bme] \cdot (\bme - \bme^*)~; ~~ \text{Cov}\big[~\bmnlMC~\big|~ \{ \Gamma_\eta \}~\big] =\frac{J[\bme^*]}{\tau_\Gamma} ~.\label{eq:logLgradMC}
\end{equation}

In this framework, the learning dynamics becomes stochastic and ruled by the master equation:
\begin{equation}
P_{n+1}( \bme')  =\!\!  \int \! d\!\! \bme~ P_n( \bme ) ~W_{\bme \to \bme'}[\bme]~; \quad W_{\bme \to \bme'}[\bme] = \text{Prob}\big( \bmnlMC = \bme' - \bme \big)  \label{eq:master},
\end{equation}
where $W_{\bme \to \bme}[\bme]$ is the probability of transition from $\bme$ to  $\bme'$.
For sufficiently large $\tau_\Gamma$ and thanks to the equalities (\ref{eq:equalities}), the central limit theorem ensures that the  unique stationary solution of (\ref{eq:master}) is a Normal Distribution with moments:
\begin{equation}
\mathbf{E}\big[~ \bme ~\big|~ P_{\infty}(\bme) ~\big] =  \bme^*~; \quad \text{\bf Cov}\big[~\bme~\big|~ P_{\infty}(\bme)~\big] = \frac{\alpha}{(2-\alpha) \tau_\Gamma} \chi^{-1}[\bme^*]~. \label{eq:stationary}
\end{equation}

\begin{algorithm}[t]
\vspace{0.2cm}
 \KwData{$\bmTO$,$\chi_\Omega$}  
 \KwResult{$\bme^*$,${\text{\bf T}}_{\eta^*}^\text{MC}$ }
 Initialization: set $\tau_\Gamma=\tau_\Omega$,$\alpha=1$ and $\bme_0$; estimate ${\text{\bf T}}_{\eta_0}^\text{MC}$ and compute $\epsilon_0$ \;
 \While{$\epsilon>1$}{

  $\bme_{n+1} \leftarrow \bme_n -  \alpha \chi_\Omega^{-1} \cdot \bmn l[\eta_n] $\;
estimate ${\text{\bf T}}_{\eta_{n+1}}^\text{MC}$ and compute  $\epsilon_{n+1}$\;
  \eIf{$\epsilon_{n+1}<\epsilon_{n} $}{
   increase $\alpha$, keeping $\alpha \leq 1$ \;
   }{
    decrease $\alpha$ and set  $\bme_{n+1} = \bme_n$\;
  }
   $n \leftarrow n+1$\;
 }
Fix $\alpha<1$ and perform several iterations.
\vspace{0.2cm}
 \caption{Algorithm pseudocode for the Ising model inference.}
\label{pseudocode}
\end{algorithm}

\textbf{Algorithm. $\quad$}
Thanks to (\ref{eq:equalities}) one may compute the mean and covariance of the model posterior distribution (with flat prior):
\begin{equation}
\mathbf{E}\big[~ \bme ~\big|~ P^\text{Post}(\bme) ~\big] =  \bme^*~; \quad \text{\bf Cov}\big[~\bme~\big|~ P^\text{Post}(\bme)~\big] = \frac{1}{\tau_\Omega} \chi^{-1}[\bme^*] \label{eq:posterior}
\end{equation}
where $\tau_\Omega$ is the size of the training dataset. 
From (\ref{eq:logLgradMC}), if $\bme \sim P^\text{Post}$ we have:
\begin{equation}
\mathbf{E}\big[~ \bmnlMC~\big|~ \{ \Gamma_{\eta \sim P^\text{Post}} \} ~\big] = 0~; ~~ \text{Cov}\big[~\bmnlMC~\big|~ \{ \Gamma_{\eta \sim P^\text{Post}} \}~\big] =\frac{2 \chi[\bme^*]}{\tau_\Gamma} ~.\label{eq:post_gradL}
\end{equation}
Interestingly, by imposing:
\begin{equation}
\frac{1}{\tau_\Omega} = \frac{\alpha}{(2-\alpha) \tau_\Gamma}\label{eq:matching}
\end{equation}
the moments (\ref{eq:stationary}) equal (\ref{eq:posterior}) \cite{Ferrari16a}.
To evaluate the inference error at each iteration we define:
\begin{equation}
\epsilon_n = \norm{ \bmnlMCt}_{\chi_\Omega} = \sqrt{\frac{\tau_\Omega}{2 D}  \bmnlMCt \cdot \chi_\Omega^{-1} \cdot \bmnlMCt}~.
\end{equation}
Averaging $\epsilon$ over the posterior distribution, see (\ref{eq:post_gradL}), gives $\epsilon = 1$.
Consequently, if $\bme_n \neq \bme^*$ implies $\epsilon_n>1$ with high probability, for $\bme_n \to \bme^*$ thanks to (\ref{eq:matching}) we expect $\epsilon_n = \sqrt{ \tau_\Omega/\tau_\Gamma/(2-\alpha)}$ \cite{Ferrari16a}.
As sketched in pseudocode \ref{pseudocode}, we iteratively update $\bme_n$ through (\ref{eq:DDupdate}) with $\tau_\Gamma=\tau_\Omega$ and $\alpha<1$ until $\epsilon_n<1$ is reached.

%% file: analysis.tex
\begin{figure}[ht]
\begin{center}
\includegraphics[clip=true,keepaspectratio,angle=-0,width=0.49\columnwidth]{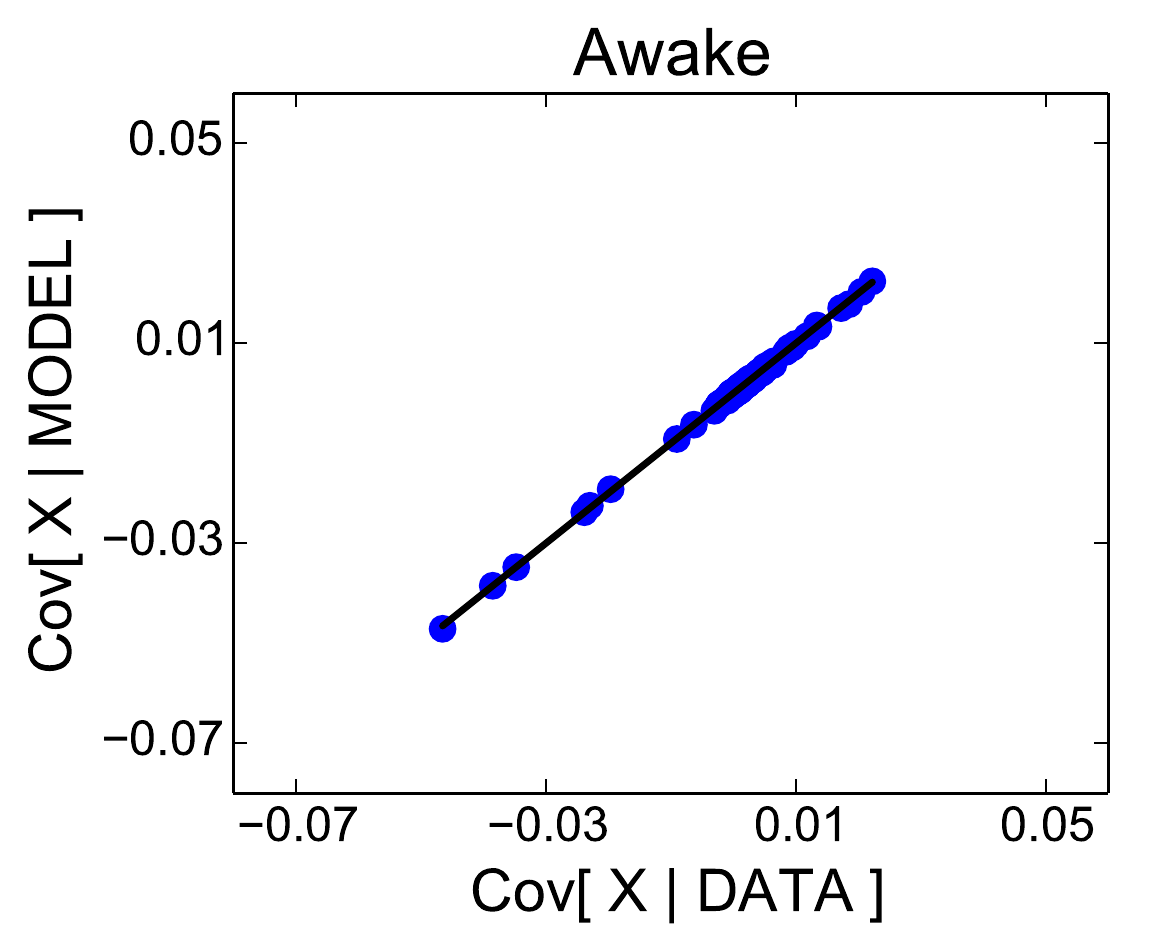}
\includegraphics[clip=true,keepaspectratio,angle=-0,width=0.49\columnwidth]{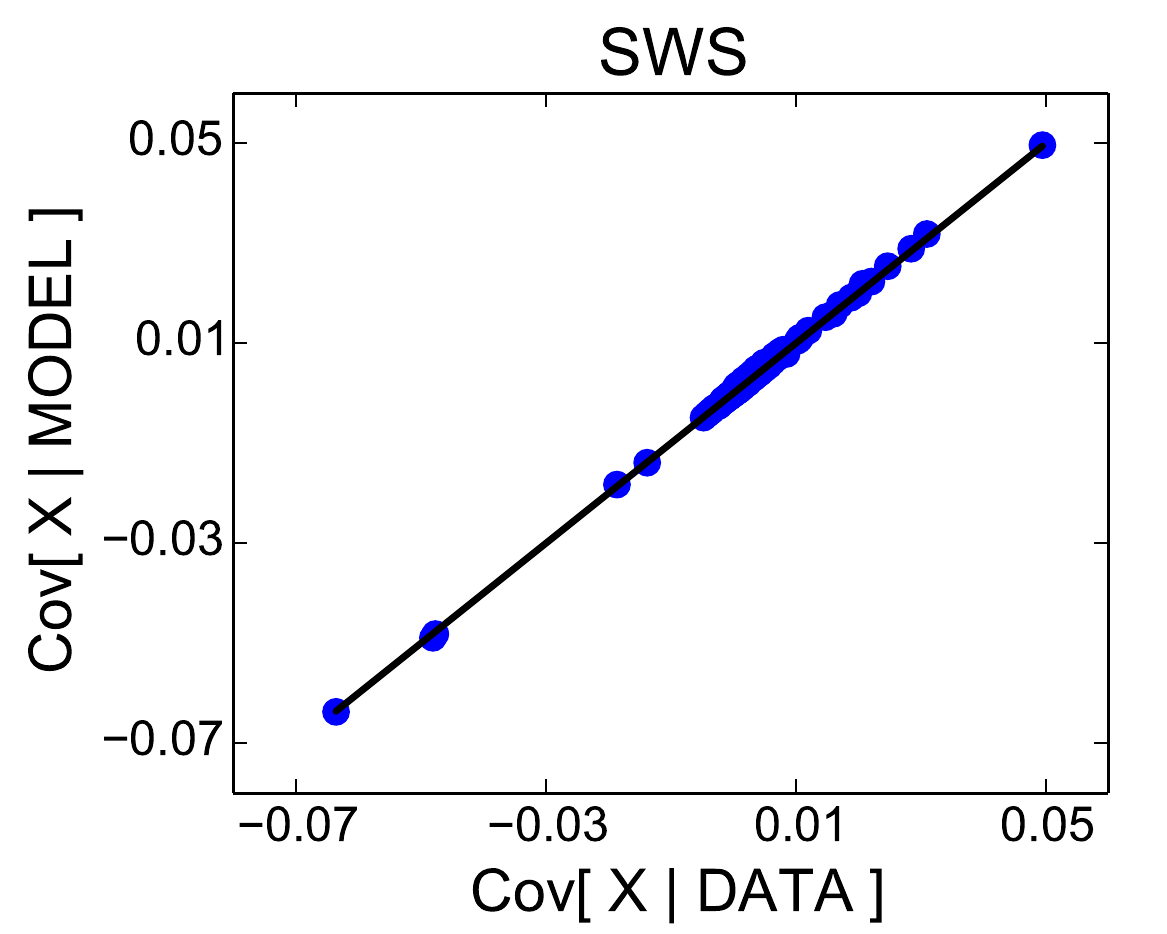}
\end{center}
\caption{Empirical pairwise covariances against their model prediction for Awake and SWS.
The goodness of the match implies that the inference task was successfully completed.
Note the larger values in SWS than Awake
}
\label{fig:corr}
\end{figure}

As in \cite{Peyrache12,Dehghani16}, we analyze $\sim 12$ hours of intracranial multi-electrode array recording of neurons in the temporal cortex of a single human patient.
The dataset is composed of the spike times of $N=59$ neurons, including $N^\text{I}=16$ inhibitory neurons and $N^\text{E}=43$ excitatory neurons.
During the recording session, the subject alternates between different brain states \cite{Peyrache12}.
Here we focused on wakefulness (Awake) and Slow-Wave Sleep (SWS) periods.
First, we divided each recording into $\tau_\Omega$ short $50ms$-long time bins and encoded the activity of each neuron $i=1,\dots, N$ in each time bin $t=1,\dots,\tau_\Omega$  as a binary variable $X_i(t) \in [0,1]$ depending on whether the cell $i$ was silent ($X_i(t)=0$) or emitted at least one spike ($X_i(t)=1$) in the time window $t$. 
We thus obtain one training dataset $\Omega = \{ \{X_i(t)\}_{i=1}^N \}_{t=1}^{\tau_\Omega}$ per brain state of interest. 
To apply the Ising model we assume that this binary representation of the spiking activity is representative of the neural dynamics and that subsequent time-bins can be considered as independent.
We then run the inference algorithm on the two datasets separately to obtain two sets of Ising model parameters $\bme_\text{Awake}$ and $\bme_\text{SWS}$.

Thanks to (\ref{eq:logLgrad}), when the log-likelihood is maximized, the pairwise Ising model reproduces the covariances $\mathbf{E}\big[~ X_i X_j~|~ \Omega~\big]$ for all  pairs $i \neq j$.
To validate the inference method, in Fig.~\ref{fig:corr} we compare the empirical and model-predicted pairwise covariances and found that the first were always accurately predicted by the second in both Awake and SWS periods.

\begin{figure}[ht]
\begin{center}
\includegraphics[clip=true,keepaspectratio,angle=-0,width=0.49\columnwidth]{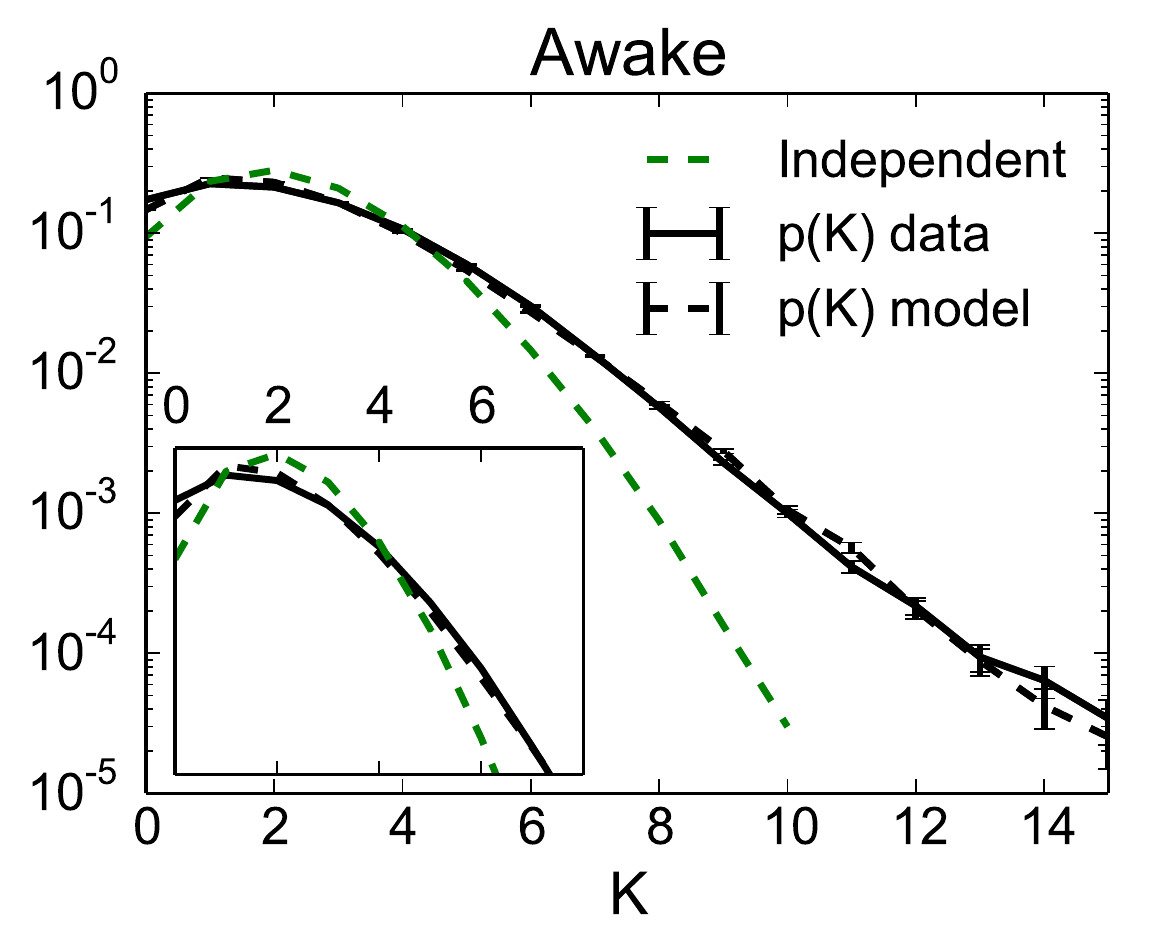}
\includegraphics[clip=true,keepaspectratio,angle=-0,width=0.49\columnwidth]{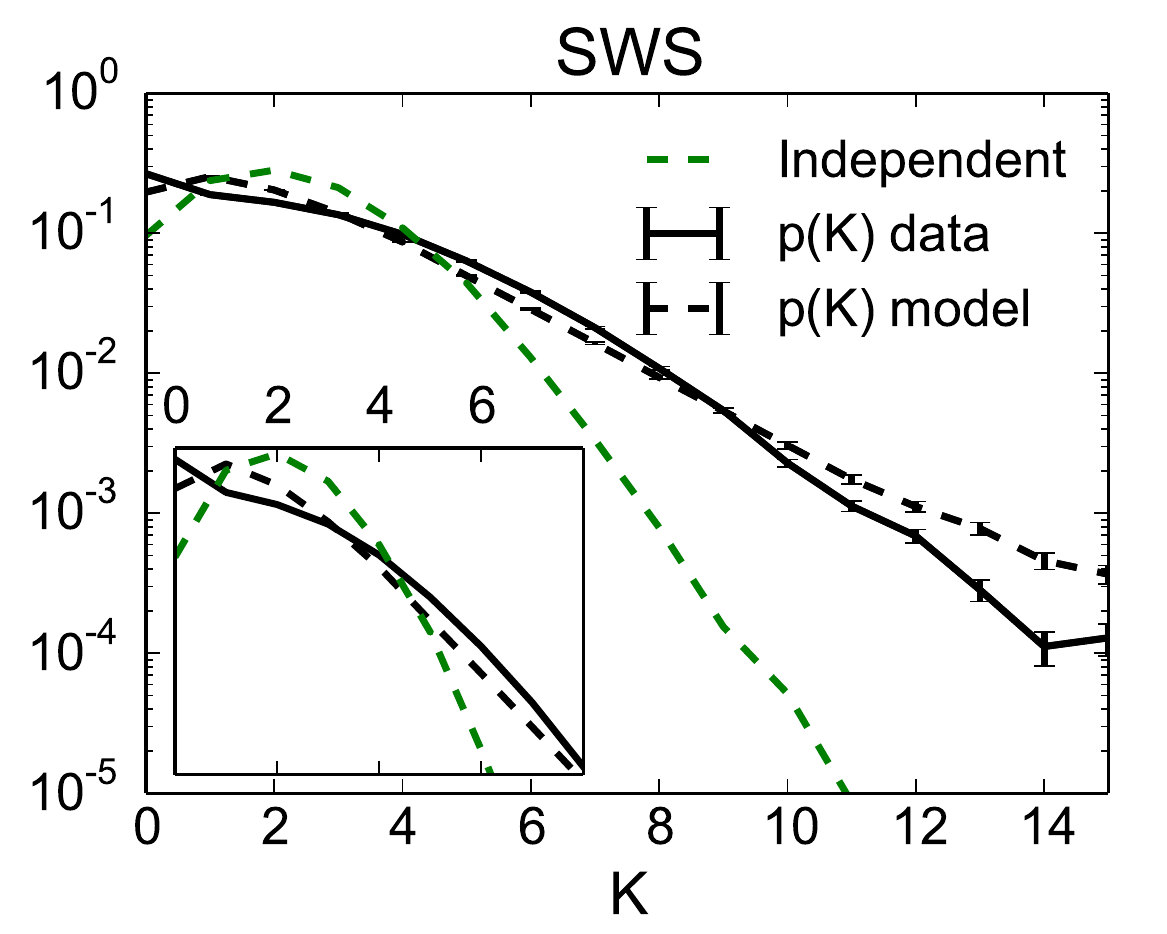}
\end{center}
\caption{Empirical and predicted distributions of the whole population activity $K = \sum_i X_i$. 
For both Awake and SWS periods the pairwise Ising model outperforms the independent model (see text).
However, Ising is more efficient at capturing the population statistics during Awake than SWS, expecially for medium and large K values.
This is consistent with the presence of transients of high activity during SWS.}
\label{fig:pOfK}
\end{figure}

This shows that the inference method is successful. 
Now we will test if this model can describe well the statistics of the population activity.
In particular, synchronous events involving many neurons may not be well accounted by the pairwise nature of the Ising model interactions.
To test this, as introduced in Ref. \cite{Tkacik14}, we quantify the empirical probability of having $K$ neurons active in the same time window: $K = \sum_i X_i$.
In Fig.~\ref{fig:pOfK} we compare empirical and model prediction for $P(K)$ alongside with the prediction from an \textit{independent} neurons model, 
the maximum entropy model that as sufficient statistics has only the single variables and not the pairwise: $\{T_a({\bm X})\}_{a=1}^N = \{X_i\}_{i=1}^N$. 
We observed that the Ising model always outperforms the independent model in predicting $P\big( ~K ~\big)$.

\begin{figure}[ht]
\begin{center}
\includegraphics[clip=true,keepaspectratio,angle=-0,width=0.49\columnwidth]{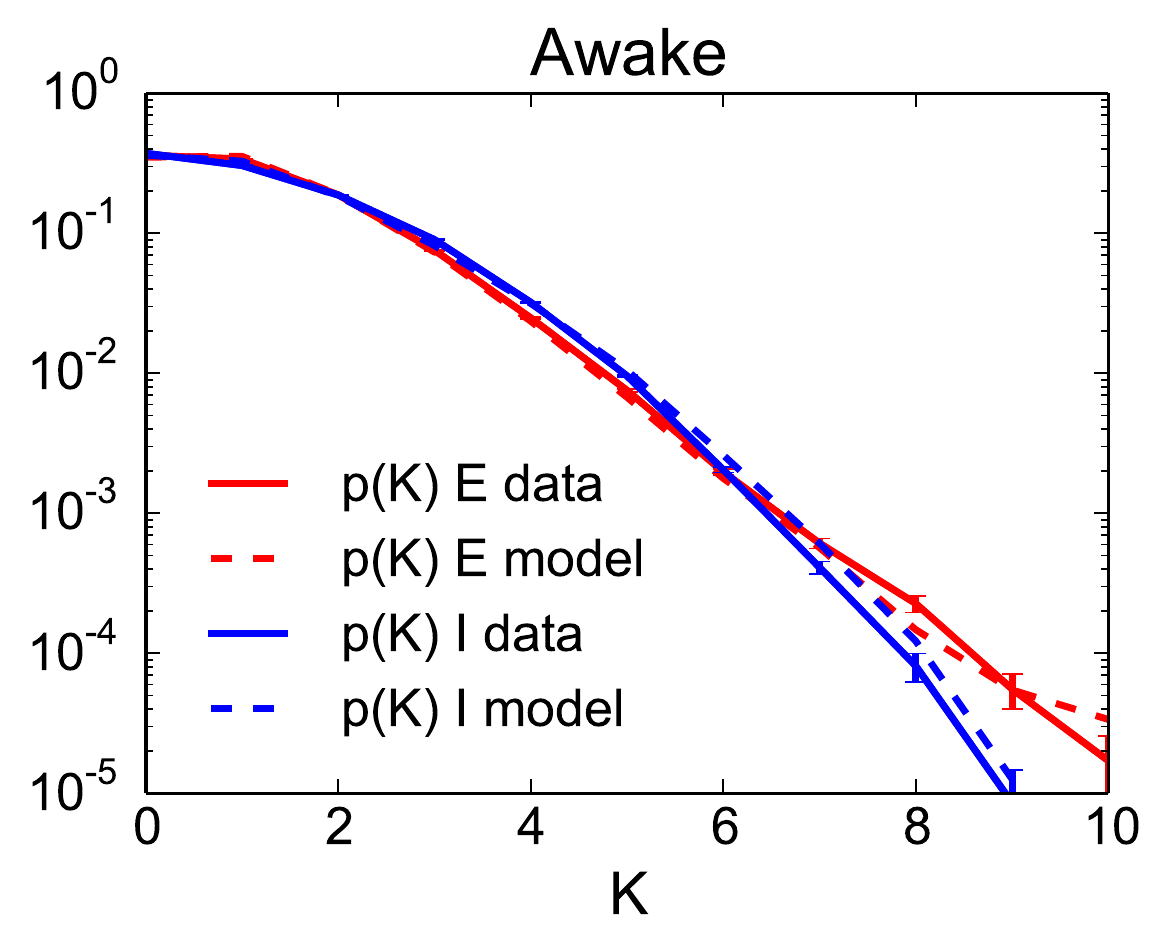}
\includegraphics[clip=true,keepaspectratio,angle=-0,width=0.49\columnwidth]{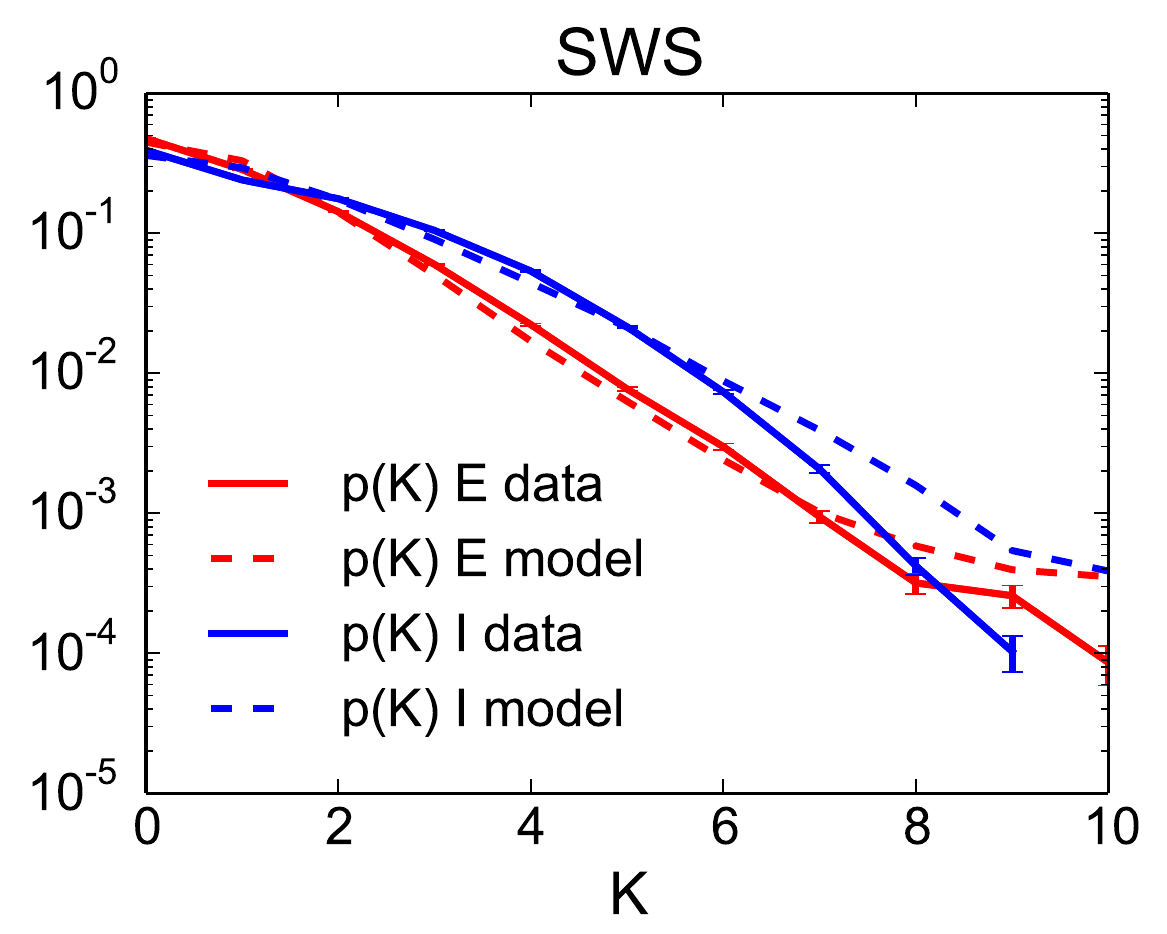}
\end{center}
\caption{Empirical and predicted distributions of excitatory (red) and inhibitory (blue) population activity.
During SWS, the pairwise Ising model fails at reproducing high activity transients, especially for inhibitory cells.}
\label{fig:pOfK_EI}
\end{figure}

Fig.~\ref{fig:pOfK} shows that the model performance are slightly better for Awake than SWS states.
This is confirmed by a Kullback-Leibler divergence estimate:
\begin{equation}
D_\text{KL}\big(~ P_\text{Awake}^\text{Data}(K)~\big|~ P_\text{Awake}^\text{Ising}(K)~\big) = 0.005; ~ D_\text{KL}\big(~ P_\text{SWS}^\text{Data}(K)~\big|~ P_\text{SWS}^\text{Ising}(K)~\big) = 0.030~. \nonumber
\end{equation} 
This effect can be ascribed to the presence of high activity transients, known to modulate neurons activity during SWS \cite{renart10} and responsible for the larger covariances, see Fig.~\ref{fig:corr} and  the heavier tail of $P(K)$, Fig.~\ref{fig:pOfK}. 
These transients are know to be related to an unbalance between the contributions of excitatory and inhibitory cells to the total population activity \cite{Dehghani16}.
To investigate the impact of these transients, in Fig.~\ref{fig:pOfK_EI} we compare $P(K)$ for the two populations with the corresponding Ising model predictions.
For the Awake state, the two contributions are very similar, probably in consequence of the excitatory/inhibitory balance \cite{Dehghani16}.
Moreover the model is able to reproduce both behaviors.
For SWS periods, instead, the two populations are less balanced \cite{Dehghani16}, with the inhibitory (blue line) showing a much heavier tail. 
Moreover, the model partially fails in reproducing this behavior, notably strongly overestimating large $K$ probabilities.

\textbf{Conclusions:}$\quad$
(i) The pairwise Ising model offers a good description of the neural network activity observed during wakefulness. 
(ii) By contrast, taking into account pairwise correlations is not sufficient to describe the statistics of the ensemble activity during SWS, where
(iii) alternating periods of high and low network activity introduce high order correlations among neurons, especially for inhibitory cells \cite{renart10}.
(iv) This suggests that neural interactions during wakefulness are more local and short-range, whereas 
(v) these in SWS are partially modulated by internally-generated activity, synchronizing  neural activity across long distances \cite{Peyrache12,renart10,LeVanQuyen16}.

%% file: ising_cortex.bbl
\begin{thebibliography}{5}

\bibitem{Ferrari16a}
U.~Ferrari.
  Learning maximum entropy models from finite-size data sets: A fast
  data-driven algorithm allows sampling from the posterior distribution.
  Phys. Rev. E, 94:023301, 2016.

\bibitem{Stevenson11}
I.H. Stevenson and K.P. Kording.
  How advances in neural recording affect data analysis.
  Nature neuroscience, 14(2):139--142, 2011.

\bibitem{Schneidman06}
E.~Schneidman, M.~Berry, R.~Segev, and W.~Bialek.
  Weak pairwise correlations imply strongly correlated network
  states in a population.
  Nature, 440:1007, 2006.

\bibitem{Peyrache12}
A.~Peyrache, N.~Dehghani, E.N. Eskandar, J.R. Madsen, W.S. Anderson, J.A.
  Donoghue, L.R. Hochberg, E.~Halgren, S.S. Cash, and A.~Destexhe.
  Spatiotemporal dynamics of neocortical excitation and inhibition
  during human sleep.
  Proceedings of the National Academy of Sciences,
  109(5):1731--1736, 2012.

\bibitem{Hamilton13}
L.~S. Hamilton, J.~Sohl-Dickstein, A.~G. Huth, V.~M. Carels, K.~Deisseroth, and
  S.~Bao.
  Optogenetic Activation of an Inhibitory Network Enhances
  Feedforward Functional Connectivity in Auditory Cortex.
  Neuron, { 80}:1066--76, 2013.

\bibitem{Tkacik14}
G.~Tkacik, O.~Marre, D.~Amodei, E.~Schneidman, W~Bialek, and Berry M.J.
  Searching for collective behaviour in a network of real neurons.
  PloS Comput. Biol., 10(1):e1003408, 2014.

\bibitem{Dehghani16}
N.~Dehghani, A.~Peyrache, B.~Telenczuk, M.~Le~Van~Quyen, E.~Halgren, S.S. Cash,
  N.G. Hatsopoulos, and A.~Destexhe.
  Dynamic balance of excitation and inhibition in human and monkey
  neocortex.
 Scientific Reports, 6(23176), 2016.

\bibitem{Gardella16}
C.~Gardella, O.~Marre, and T.~Mora.
  A tractable method for describing complex couplings between neurons
  and population rate.
  eneuro, 3(4):0160, 2016.

\bibitem{Tavoni17}
G.~Tavoni, U.~Ferrari, S.~Cocco, F.P. Battaglia, and R.~Monasson.
  Functional coupling networks inferred from prefrontal cortex
  activity show experience-related effective plasticity.
  Network Neuroscience, MIT press, 2017.

\bibitem{Jaynes82}
E.T. Jaynes.
 On The Rationale of Maximum-Entropy Method.
 Proc. IEEE, 70:939, 1982.

\bibitem{Ferrari17}
U.~Ferrari, T.~Obuchi, and T.~Mora.
  Random versus maximum entropy models of neural population activity.
  Phys. Rev. E, 95:042321, 2017.

\bibitem{Ackley85}
D.~H. Ackley, G.~E. Hinton, and T.~J. Sejnowski.
  A learning algorithm for boltzmann machines.
  Cognitive Science, 9:147--169, 1985.

\bibitem{Broderik07}
T.~Broderick, M.~Dudik, G.~Tkacik, R.E. Schapire, and W.~Bialek.
  Faster solutions to the inverse pairwise Ising problem.
  Arxiv:0712.2437, 2007.

\bibitem{Cocco11}
S.~Cocco and R.~Monasson.
  Adaptive cluster expansion for inferring Boltzmann machines with
  noisy data.
  Phys. Rev. Lett., { 106}:090601, 2011.

\bibitem{Sohl-Dickstein11}
J.~Sohl-Dickstein, P.~B. Battaglino, and M.~R. DeWeese.
  New Method for Parameter Estimation in Probabilistic Models:
  Minimum Probability Flow.
  Phys. Rev. Lett., { 107}:220601, 2011.

\bibitem{renart10}
A.~Renart, J.~De~La~Rocha, P.~Bartho, L.~Hollender, N.~Parga, A.~Reyes, and
  K.D. Harris.
  The asynchronous state in cortical circuits.
  Science, 327(5965):587--590, 2010.

\bibitem{Amari98}
S.~Amari.
  Natural Gradient Works Efficiently in Learning, Neural
  Computation.
  Neural Comput., 10:251--276, 1998.

\bibitem{Amari98b}
S.~Amari and S.C. Douglas.
Why natural gradient?
 Proc. IEEE, 2:1213--16, 1998.
  
 \bibitem{LeVanQuyen16}
 M.~Le Van Quyen,L.E.~Muller, B.~Telenczuk, E.~Halgren, S.~Cash, N.G.~Hatsopoulos, N.~Dehghani, and A.~Destexhe
 High-frequency oscillations in human and monkey neocortex during the wake--sleep cycle.
 Proceedings of the National Academy of Sciences, 113, 33, 9363--68, 2016
 
\end{thebibliography}
